\documentclass[fleqn,usenatbib]{mnras}

\usepackage{newtxtext}
\usepackage{mathptmx}
\usepackage{txfonts}

\usepackage[T1]{fontenc}

\DeclareRobustCommand{\VAN}[3]{#2}
\let\VANthebibliography\thebibliography
\def\thebibliography{\DeclareRobustCommand{\VAN}[3]{##3}\VANthebibliography}

\usepackage{graphicx}	% Including figure files

\usepackage{calrsfs}
\DeclareMathAlphabet{\mathdutchcal}{U}{dutchcal}{m}{n}

\title[Universality of the r-process abundances]{``Nuclear thermometers'' reveal the origin of the universal r-process nucleosynthesis}

\author[Jos\'e Nicol\'as Orce et al.]{Jos\'e Nicol\'as Orce,$^{1,2}$\thanks{Email:jnorce@uwc.ac.za;
\linebreak https://github.com/UWCNuclear;   http://nuclear.uwc.ac.za}
\\
$^{1}$Department of Physics \& Astronomy, University of the Western Cape, P/B X17, Bellville 7535, South Africa\\
$^{2}$National Institute for Theoretical and Computational Sciences (NITheCS), South Africa\\
}

\date{Accepted XXX. Received YYY; in original form ZZZ}

\pubyear{2024}

\begin{document}
\label{firstpage}
\pagerange{\pageref{firstpage}--\pageref{lastpage}}
\maketitle

\begin{abstract}

The resembling behaviour of giant dipole resonances built on ground and excited states supports the validity of the Brink-Axel hypothesis and
% , which present similar energy systematics. Together with previous work,
assigns giant dipole resonances as spectroscopic probes --- or ``nuclear thermometers'' ---
to explore the cooling of the kilonova ejecta in neutron-star mergers down to the production
% of about 50\%
of  heavy elements beyond iron through the rapid-neutron capture or r-process.
In previous work, we found a slight energy increase in the giant dipole resonance built on excited states at the typical temperatures of $1.0\gtrapprox T\gtrapprox0.7$ MeV where seed nuclei are produced, before ongoing neutron capture.
% which enhances the symmetry energy in the well-known Bethe-Weizs\"acker semi-empirical mass formula and lowers the nuclear binding energy.
% However, our previous work did not reach the  temperatures of $T\lesssim0.5$ MeV (or $6\times 10^9$ K)
% where the r-process is expected to create about 50\% of the heavy elements beyond iron.
Crucial data  are presented here  supporting an enhanced symmetry energy at $T=0.51$ MeV (or $5.9\times 10^9$ K)
--- where the r-process occurs --- that lowers the binding energy in the Bethe-Weizs\"acker semi-empirical mass formula
and results in the close in of the neutron drip line. Ergo, providing an origin to  the universality
of elemental abundances by limiting the reaction network for r-process nucleosynthesis.
An enhanced symmetry energy away from the ground state is further supported  by shell-model calculations of the nuclear
electric dipole ({\sc E1}) polarizability --- inversely proportional to the symmetry energy ---
as a result of the destructive contribution of the products of off-diagonal {\sc E1} matrix elements.
% Further experimental work on neutron-rich nuclei at  $T\lesssim0.5$ MeV is crucial.

\end{abstract}

% Select between one and six entries from the list of approved keywords.
% Don't make up new ones.

\maketitle

\section{Motivation}
\label{intro}

Astronomical observations indicate that the abundances of heavy elements from barium to lead in metal-poor stars are consistent with the scaled Solar system abundance pattern
for the rapid-neutron capture or r-process. About 50\% of the heavy elements beyond iron are produced through the r-process during the cooling down of the kilonova ejecta that follows from
neutron star mergers~\citep{watson2019identification,sneppen2023discovery};  other potential r-process scenarios are supernova~\citep{reichert2021nucleosynthesis} and
collapsars~\citep{siegel2019collapsars}. Given that the Sun formed billions of years after these metal-poor stars, from gas that was enriched by many stellar generations in various ways, such an astounding agreement suggests that the way these elements are produced is universal~\citep{frebel2018nuclei,sneden2008neutron,ji2016r}.
A detailed interpretation for the universality of elemental abundances has long been obscured with a couple of scenarios being proposed: 1) an artifact of nuclear properties such as binding energies and $\beta$-decay rates;
2) a single cosmic site with astrophysical conditions that are generated uniformly throughout cosmic time~\citep{kajino2019current}. The abundance of heavy elements is intrinsically linked to the limits of nuclear stability, the so-called  neutron drip line, after which nuclei become unbound.

The exact whereabouts of the neutron drip line for heavy nuclei remain unknown because of the lack of experimental information~\citep{wang2015positioning} --- mainly atomic masses ---
in the so-called \emph{terra incognita} of exotic nuclei~\citep{chomaz2003introduction}. There are two major unsettling scenarios that may critically affect the limits of bound nuclei
and sensitivity studies of r-process abundances: 1) Atomic masses cannot be measured for most of the exotic nuclei,
critically affecting mass-model predictions,  which deviate from each other substantially  --- a few MeVs as nuclei become more neutron rich~\citep{mumpower100keV} ---
and sensitivity studies of r-process abundances~\citep{mumpower2016impact}. The latter may need mass models with rms deviations of less than 100 keV with respect to the available mass data
to be able to distinguish between different astrophysical trajectories~\citep{mumpower100keV}. A few mass models present rms deviations of about 0.3 MeV, with most in the range of 0.5 MeV;
for instance, a mass uncertainty of $\pm$0.5 MeV may lead to an abundance uncertainty of a factor around 2.5~\citep{jiang2021sensitivity}.
2) Available mass data are generally for nuclei in their ground states~\citep{wang2021ame} --- with a few isomeric ones --- and sensitivity studies of r-process abundances
assume that masses remain unchanged at the high temperatures involved in stellar explosions. Although this may not necessarily be the case~\citep{donati1994temperature}.

A  solution to this fundamental question has recently been suggested~\citep{orce2023enhanced} as arising from the structure and dynamics of the giant dipole resonance ({\sc GDR}) as a function of temperature~\citep{snover1986giant,gaardhoje1992nuclear,schiller2007compilation}. Within the hydrodynamic model, the {\sc GDR} is a collective motion of protons and neutrons out of phase
characterized by the nuclear symmetry energy $a_{sym}$~\citep{berman1975measurements,danos1958long}, i.e.,
the coefficient of the $(N-Z)^2/A$ term in the semi-empirical mass formula~\citep{weizsacker1935theorie,bethe1936nuclear}. The {\sc GDR} is
responsible for most of the absorption and emission of electromagnetic radiation in nuclei.
Microscopically, the {\sc GDR}  can complementary be described as the superposition of 1$\hbar\omega$ particle-hole excitations within the shell model~\citep{balashov1962relation,orce2023global,orce2023electric}.
Using Danos' second-sound hydrodynamic model~\citep{danos1958long},
the centroid energy ($E_{_{GDR}}$) and width ($\Gamma_{_{GDR}}$) of the {\sc GDR} are related to $a_{sym}$ by,
% a modification of the Steinwedel-Jensen ({\small SJ}) model~\citep{steinwedel1950nuclear} in
\begin{equation}
E_{_{GDR}} = \frac{\hbar\mathdutchcal{k}}{A} \left( \frac{8  NZ a_{sym}}{M} \left[ 1 - \left( \frac{ \Gamma_{_{GDR}} }{2E_{_{GDR}}} \right)^2 \right]  ~\right)^{1/2},
\label{eq:EGK}
\end{equation}
where $A=N+Z$ is the atomic mass number,  $M=931.494$ MeV/c$^2$ the nucleon mass and  $\mathdutchcal{k}$ the real eigenvalue of the Helmholtz equation $\bigtriangledown^2\rho_{_Z}+\mathdutchcal{k}^2\rho_{_Z}=0$, with the
boundary condition $(\mathbf{\hat{n}} \bigtriangledown\rho_{_Z})_{_{surface}}=0$, and has a value of
$\mathdutchcal{k}R=2.082$ for a spherical nucleus~\citep{rayleigh1896theory}, with $R=1.2 A^{1/3}$.
Trivially~\citep{bergere1977lecture},
\begin{equation}
 a_{sym}(A) = 9.937 \times 10^{-4}  ~ \bigg(\frac{A^{8/3}}{NZ}\bigg) ~\frac{ E_{_{_{_{GDR}}}}^2}{1-\bigg(\frac{\Gamma_{_{GDR}}}{2 E_{_{_{_{GDR}}}}} \bigg)^2}.
\label{GDRsymm}
\end{equation}

These equations were derived for {\sc GDR}s built on ground states but one could also  apply them to {\sc GDR}s built on excited states
% if the
% Brink-Axel hypothesis holds at the temperatures {\sc T} occurring during the creation of elements.
assuming the validity of the Brink-Axel hypothesis, i.e., that a {\small GDR} can be built on every state in a nucleus~\citep{brink1955some,axel1962electric}.
This opens the exciting prospect of employing {\sc GDR}s built on excited states as \emph{``nuclear thermometers''} in order to investigate the cooling down of the
ejecta gas following stellar explosions~\citep{orce2023enhanced}. A similar strategy led to the findings of shell effects at high-excitation energies through dips of
the photo-absorption cross section in semimagic nuclei~\citep{ngwetsheni2019continuing,ngwetsheni2019combined,ngwetsheni2019we}.
% as presented at the IV International Conference on Nuclear Structure and Dynamics (NSD2019) in Venice (Italy)~\citep{ngwetsheni2019we}.
Additional support of the Brink-Axel hypothesis comes from experimental data~\citep{snover1986giant,gaardhoje1992nuclear,guttormsen2016validity}
and theoretical calculations~\citep{sagawa1984self,sommermann1983microscopic,ring1984microscopic,jang1985two}.
Namely, {\small GDRs} built on excited states below the critical temperatures and spins where the {\small GDR} width starts broadening
--- i.e., for moderate average temperatures of $T\lessapprox T_c = 0.7 +37.5/A$ MeV and spins $I$ below the critical angular momentum $I \lessapprox I_c=0.6 A^{5/6}$~\citep{kusnezov1998scaling} --- present
% , which present similar properties
similar centroid energies,
$E_{_{GDR}}$, and resonance  strengths, $S_{_{GDR}}$ --- relative to the  classical dipole oscillator strength or Thomas--Reiche--Kuhn ({\sc TRK})
sum rule~\citep{ladenburg1923absorption,thomas1925zahl,reiche1925zahl,kuhn1925gesamtstarke,levinger1960nuclear,bohr1998nuclear} --- to those found for the ground-state counterparts~\citep{snover1986giant,gaardhoje1992nuclear,bortignon2019giant}, which indicate a common physical origin for all {\sc GDR}s~\citep{brink1955some,axel1962electric}.
% , in concordance with the Brink--Axel hypothesis
Deviations from the Brink--Axel hypothesis have nonetheless been indicated at low excitation energies, in the pygmy-dipole resonance region~\citep{sieja2023brink}.

% \begin{figure}[!ht]
% % Use the relevant command for your figure-insertion program
% % to insert the figure file.
% \centering
% \sidecaption
% \includegraphics[width=12cm,height=7cm,clip]{T.eps}
% \caption{Cooling down of neutron star mergers: from kilonova to ground state (T=0 MeV).}
% \label{fig:T}       % Give a unique label
% \end{figure}

The validity of  the Brink--Axel hypothesis has been further investigated using the most recent evaluation of {\sc GDR}s built on excited states by
Schiller and Thoennessen~\citep{schiller2007compilation}.
% For deformed nuclei, we estimate a similar equation to Eq. \ref{GDRsymm}, but using the average centroid energy and the
% {\small FWHM} of the total Lorentzian (see e.g.~\citep{gaardhoje1988nuclear}). Uncertainties in the quoted values arise from the error propagation of Eq.~\ref{GDRsymm}.
For quadrupole deformed nuclei with an eccentricity of $a^2-b^2=\epsilon R^2$, where
$a$ and $b$ are the half axes and $\epsilon$ the deformation parameter, the {\small GDR} lineshape splits into
two peaks with similar values of $Ka$ and $Kb\approx2.08$~\citep{danos1958long}.
For consistency with the choice of Berman and Fultz~\citep{berman1975measurements}, the energy of the {\sc GDR} built on excited states for deformed nuclei has been
determined from the two sets of Lorentz parameters $E_1$ and $E_2$ given in the online supplementary table~\citep{schiller2007compilation}.
\begin{eqnarray}
E_{_{GDR}} &=& \frac{\left( E_1 + 2 E_2 \right)}{3}; ~\mbox{for prolate nuclei} \label{eq:pro}
\\
E_{_{GDR}} &=& \frac{\left( 2E_1 + E_2 \right)}{3};  ~\mbox{for oblate nuclei},
\label{eq:ob}
\end{eqnarray}
where the second peak is stronger for prolate deformations and viceversa for oblate ones~\citep{danos1958long,bergere1977lecture}.
Similar results are obtained using the cross section for the respective Lorentzians, $\sigma_{_1}$ and $\sigma_{_2}$ ~\citep{berman1975measurements},
\begin{equation}
E_{_{GDR}} =  E_1 \left(\frac{\sigma_{_1}}{\sigma_{_1} + \sigma_{_2}}\right) + E_2  \left(\frac{\sigma_{_2}}{\sigma_{_1} + \sigma_{_2}} \right).
\end{equation}
% or  the strengths of each of the {\sc GDR} components, $S_1$ and $S_2$ --- i.e.,  the fraction of the classical oscillator strength or {\sc TRK} sum rule~\citep{thomas1925zahl,ladenburg1923absorption,reiche1925zahl,kuhn1925gesamtstarke,bohr1998nuclear} --- provided in Ref.~\citep{schiller2007compilation}.

\begin{figure}
% Use the relevant command for your figure-insertion program
% to insert the figure file.
\centering
% \sidecaption
\includegraphics[width=8.cm,height=6.5cm,clip]{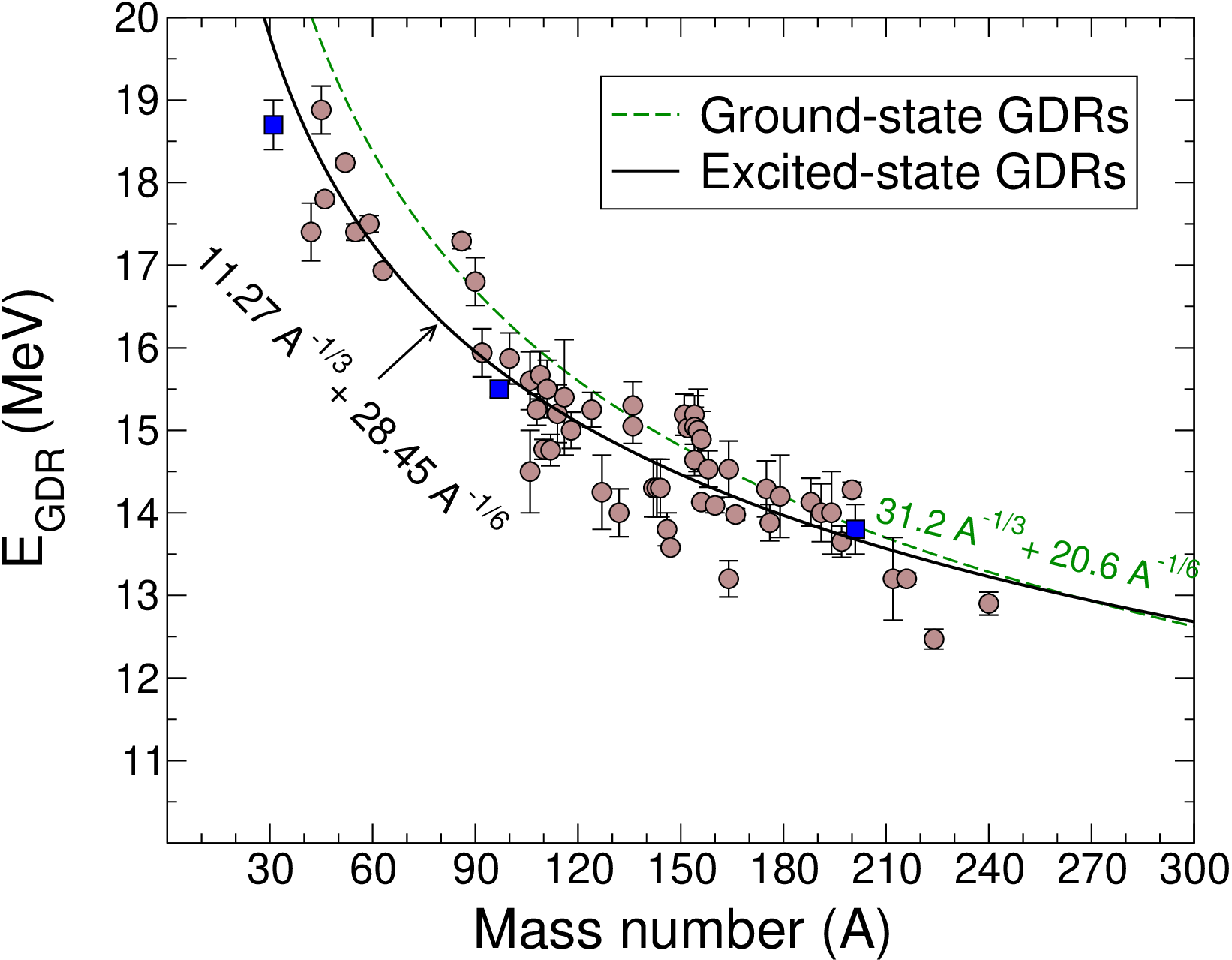}
\caption{Systematics of centroid energies for excited-state/hot {\sc GDR}s extracted from the 2007 evaluation~\citep{schiller2007compilation} and recent measurements~\citep{pandit2012critical,dey2014probing,mondal2018study,pandit2021puzzle}. The solid curve corresponds to the new parametrization, $E_{_{GDR}}=11.27 ~A^{-1/3} + 28.45~ A^{-1/6}$, proposed for hot {\sc GDR}s, whereas the dash curve is the well-known parametrization, $E_{_{GDR}}=31.2 ~A^{-1/3} + 20.6~ A^{-1/6}$, for ground states~\citep{berman1975measurements}.}
\label{fig:Egdr}       % Give a unique label
\end{figure}

Figure ~\ref{fig:Egdr} shows the $E_{_{GDR}}$ systematics  for excited-state/hot {\sc GDR}s considering the weighted average of various measurements~\citep{schiller2007compilation}
and Eqs.~\ref{eq:pro} and \ref{eq:ob} for deformed nuclei. These data yield a new parametrization of $E_{_{GDR}}=11.27 ~A^{-1/3} + 28.45~ A^{-1/6}$ with an rms relative error of 0.035,
which converges
with the ground-state parametrization, $E_{_{GDR}}=31.2 ~A^{-1/3} + 20.6~ A^{-1/6}$~\citep{berman1975measurements}, for heavy nuclei.
The discrepancy for light nuclei may arise from the splitting of the {\sc GDR}~\citep{orce2023global} since fusion-evaporation reactions do not populate
the  $T+1$ isospin component of the {\sc GDR}~\citep{gaardhoje1992nuclear,fallieros1970isovector,shoda1979isospin}, energetically favouring the $(\gamma,n)$ over the $(\gamma,p)$ decay~\citep{shoda1979isospin,shoda1975gamma,mcneill1993isospin,orce2022competition}.

Compared with the previous parametrization for {\sc GDR}s built on excited states, $E_{_{GDR}}=18 ~A^{-1/3} + 25~ A^{-1/6}$ ~\citep{gaardhoje1992nuclear},
there is a smaller weight of the $A^{-1/3}$ factor arising from the
Steinwedel-Jensen model~\citep{steinwedel1950nuclear} as compared to $A^{-1/6}$ from the Goldhaber-Teller model~\citep{goldhaber1948nuclear}.
The former considers protons and neutrons as compressible interpenetrating fluids moving within the rigid surface of the initial nucleus, with the nuclear surface area  being the
restoring force of the {\sc GDR}~\citep{berman1975measurements,bergere1977lecture},
% with the nuclear surface being the restoring force of the {\sc GDR}
% ~\citep{steinwedel1950nuclear},
whereas the latter assumes that the
proton sphere moves as a whole with respect to the neutron sphere, with a resulting displacement of the two spheres~\citep{goldhaber1948nuclear}
and the restoring force being proportional to the density gradient of the two fluids~\citep{berman1975measurements,bergere1977lecture}.

\begin{figure}
\begin{center}
% Use the relevant command for your figure-insertion program
% to insert the figure file.
\centering
\includegraphics[width=7.5cm,height=6.5cm,clip]{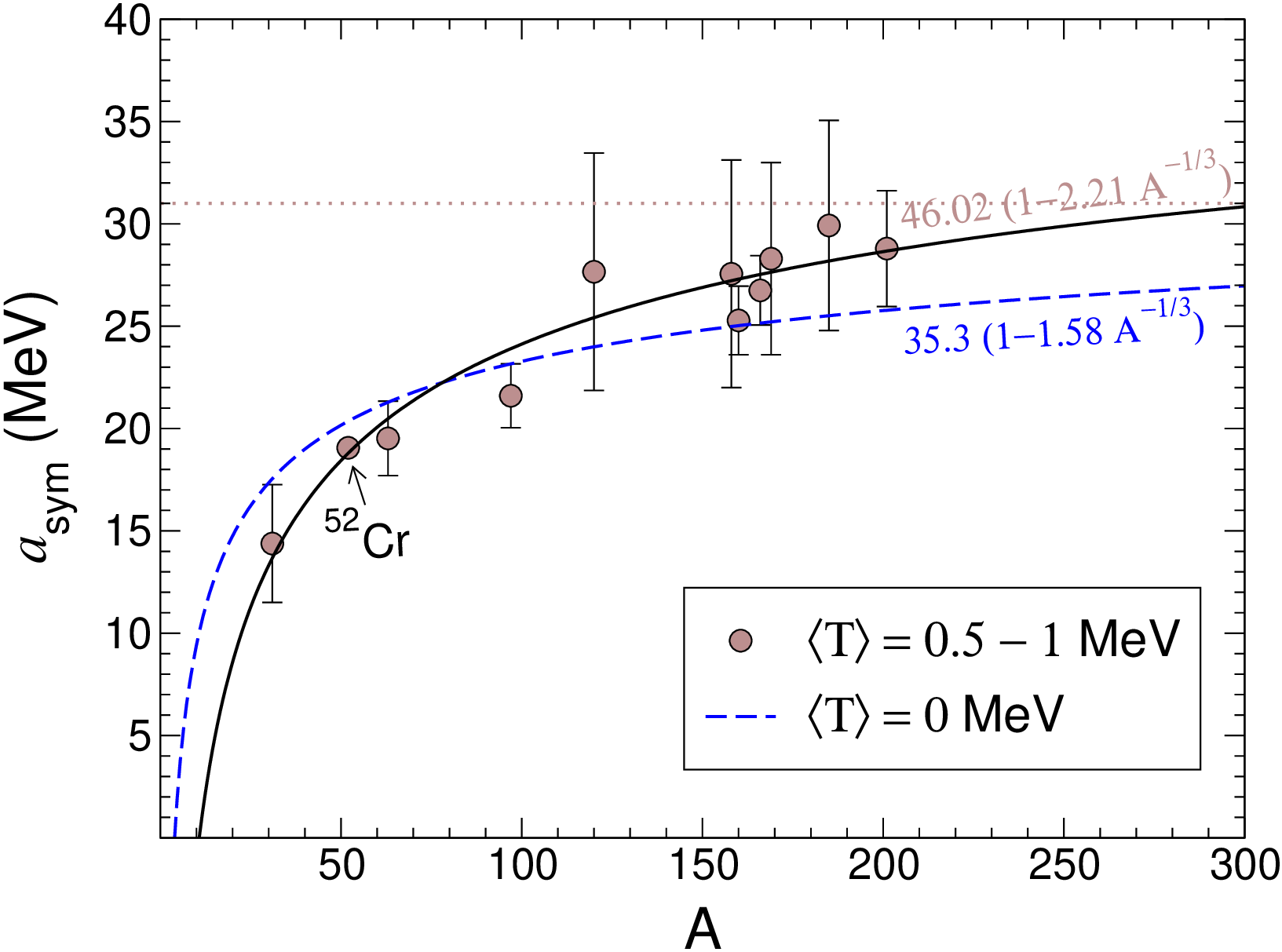}
\caption{Symmetry energy coefficient, $a_{sym}(A)$, of finite nuclei  as a function of mass number $A$ extracted using Eq.~\ref{GDRsymm} from {\small GDRs} built on ground states ($T=0$) (solid curve) and  excited states ($T=0.5-1$ MeV) (circles), including the new data point at $T=0.51$ MeV ($^{52}$Cr)~\citep{feldman1993statistical,schiller2007compilation}.}
\label{fig:GDRT}       % Give a unique label
\end{center}
\end{figure}

\section{Results}
\label{sec-1}

A more detailed analysis uncovers subtle differences with a slight energy increase of  $3-5$\%  for {\sc GDR}s built on excited states
at effective temperatures of $\langle T \rangle \approx0.7-1.0$ MeV as compared with ground-state values~\citep{orce2023enhanced}.
% for nearly-spherical nuclei.
% $^{120}$Sn~\citep{heckman2003low}, $^{208}$Pb~\citep{baumann1998evolution} and $^{201}$Tl~\citep{pandit2012critical} nuclei as well as for the deformed nuclei in the
% $A=160-180$ mass region~\citep{gaardhoje1988nuclear,gossett1985deformation,pandit2021puzzle}.
Although such an increase is within the experimental errors, it leads to the distinct systematic behaviour shown by the solid line in Fig.~\ref{fig:GDRT},
which saturates at $a_{_{sym}}\approx 31$ MeV as compared with $a_{_{sym}}\approx 27$ MeV for $\langle T \rangle \approx 0$ (dashed line).

Our original work was not be sensitive to the lower temperatures occurring during neutron capture in neutron-star mergers,
which likely range from $T\approx~0.5 \times 10^8$ K~\citep{goriely2011r} to $T\approx 6 \times 10^9$ K~\citep{wu2016production}
(i.e., in the range from $T\approx0.04$ to 0.5 MeV, respectively).

\begin{figure}
% Use the relevant command for your figure-insertion program
% to insert the figure file.
	\begin{center}
% 	\sidecaption
\includegraphics[width=8.cm,height=6.5cm,clip]{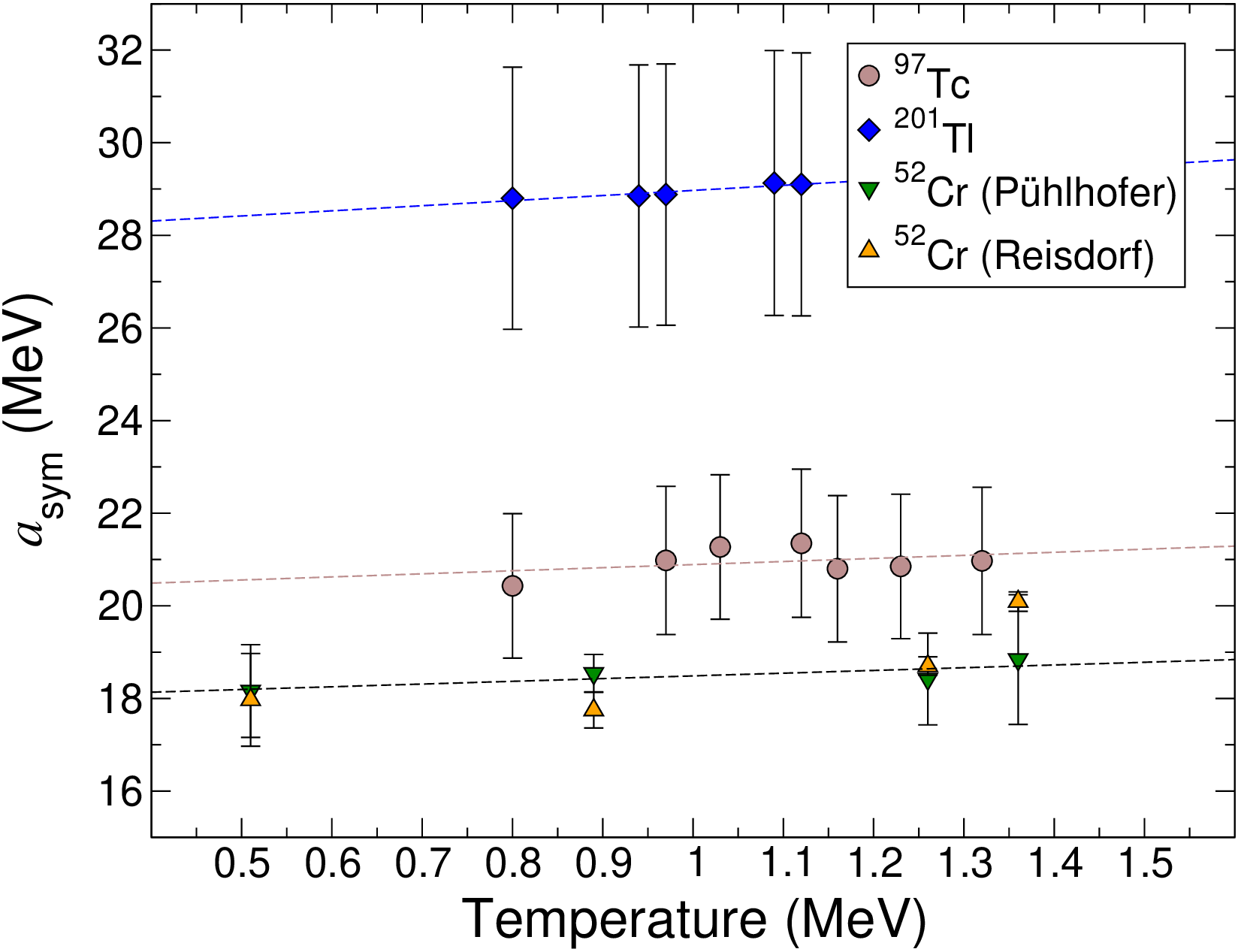}
\caption{Symmetry energy coefficient, $a_{sym}(A)$ extracted for $^{201}$Tl,  $^{97}$Tc and $^{52}$Cr as a function of temperature, T.
A similar smooth behaviour is observed for other nuclei. Dashed lines are linear regressions to the data. Two level densities approaches (P\"uhlhofer~\citep{puhlhofer1977interpretation} and Reisdorf~\citep{reisdorf1981analysis}) were applied to evaluate the effective temperature,
yielding similar results.}
\label{fig:tchange}       % Give a unique label
	\end{center}
\end{figure}

Nevertheless, Fig.~\ref{fig:tchange} shows that the symmetry energy does not change with temperature in the [0.74,1.3] MeV range, which suggests that this relation could still hold at lower temperatures down to the lower limits for the waiting point approximation, i.e., for $T\gtrapprox 1$ GK or $T\gtrapprox 0.1$ MeV, where neutron captures become balanced by high-energy photons from neutron photodisintegrations~\citep{cameron1983waiting}.

Figure~\ref{fig:GDRT} shows the symmetry energy coefficient deduced from Eq.~\ref{GDRsymm} for {\sc GDR}s built on nuclear ground states (solid curve for $T=0$ MeV)
and excited states ($T=0.5-1$ MeV) at the relevant temperatures for the creation of elements in the kilonova ejecta. More detailed explanations are available
in our previous work, where higher temperatures of $T\geq 0.74$ MeV were considered~\citep{orce2023enhanced}. Here, we focus on the new data point at $T=0.51$ MeV provided by the
$\alpha +  ^{48}$Ti $\rightarrow ^{52}$Cr fusion-evaporation reaction populated at moderate temperature and spin~\citep{feldman1993statistical,schiller2007compilation}. Results show that $E_{_{GDR}}$ is roughly constant over the range of excitation energies
$E_x=[20.4 - 35.3]$ MeV, with a slight increase of $\Gamma_{_{GDR}}$ with increasing bombarding energy. This excitation energies correspond to effective nuclear temperatures
of the states populated by $\gamma$ ray of $\langle T \rangle = [0.51 -1.36]$ MeV~\citep{feldman1993statistical}, which depend on the choice of the
level density parameters~\citep{orce2022competition}.
% Other prescriptions are available (see e.g.~\citep{orce2022competition} and references therein).

The change of $a_{sym}(A)$ as a function of $T$ for $^{52}$Cr is shown in Fig.~\ref{fig:tchange} together with other nuclei presented in Ref.~\citep{orce2023enhanced}.
Two different  prescriptions for the level density provided similar results~\citep{puhlhofer1977interpretation,reisdorf1981analysis}.
Linear regressions to the data show similar smooth decreasing trends as $T$ goes down.
The weighted average of the eight different measurements in $^{52}$Cr~\citep{feldman1993statistical} listed in the 2007  evaluation by Schiller and Thoennessen~\citep{schiller2007compilation} yields $a_{sym}(A)=19.06(13)$ MeV, following exactly the symmetry energy trend in Fig.~\ref{fig:GDRT}, which presents a similar leptodermus approximation~\citep{myers1969average} (dashed curve) to the one presented by ~\citep{orce2023enhanced}, arising from nuclear surface effects being localized within the liquid-drop model~\citep{moller2019nuclear}.

\section{Discussion and Conclusions}

A plausible explanation  is provided by Donati and co-workers, who investigated the temperature dependence of the effective nucleon mass~\citep{donati1994temperature}
--- the so-called ‘$w$’ mass~\citep{mahaux1985dynamics} used to describe the highly dynamic and, hence, nonlocal mean field in both spatial coordinates and time ---
in medium and heavy mass nuclei within the quasiparticle random phase approximation ({\sc QRPA})~\citep{bohr1998nuclear}.
% within the liquid-drop and Fermi gas models~\citep{bortignon2019giant}
% --- the so-called ‘$w$’ mass~\citep{mahaux1985dynamics} --- to account for the non-locality of the Hartree-Fock potential in spatial coordinates due to the Pauli principle.
% Such an effective nucleon mass influences the temperature dependence of the symmetry energy.
It was found that $w$ decreases as $T$ increases in the temperature interval $0<T<1$ MeV giving rise to a larger symmetry energy of approximately 8\% at $T\approx 1$ MeV~\citep{donati1994temperature} ,
which results in a larger centroid energy of the {\small GDR}, as $\hbar \omega \propto a_{_{sym}}(T)^{1/2}$.
The temperature dependence of $w$ additionally affects supernova explosions yielding a stronger shock wave~\citep{donati1994temperature}
by quenching electron capture on protons, i.e., the neutronization processes in the initial phase of the star collapse.
Support for a decreasing effective nucleon mass have recently been provided by the density functional theory ({\sc DFT}) approach extending the Skyrme interaction in order to improve the density of states around the Fermi surface~\citep{fantina2011nuclear} and $\beta$-decay studies~\citep{severyukhin2015sensitivity}.
It should be noted, however, that Monte Carlo shell-model calculations do not predict such a generic increase of the symmetry-energy coefficient as a function of temperature for $T < 1$ MeV~\citep{dean1995temperature,dean2002temperature}; a discrepancy that has been associated to the lack of correlations in the {\sc QRPA} (e.g., pairing, two-particle–two-hole, and higher-order correlations).

\begin{figure}
% Use the relevant command for your figure-insertion program
% to insert the figure file.
\centering
% \sidecaption
\includegraphics[width=8.cm,height=6.7cm,clip]{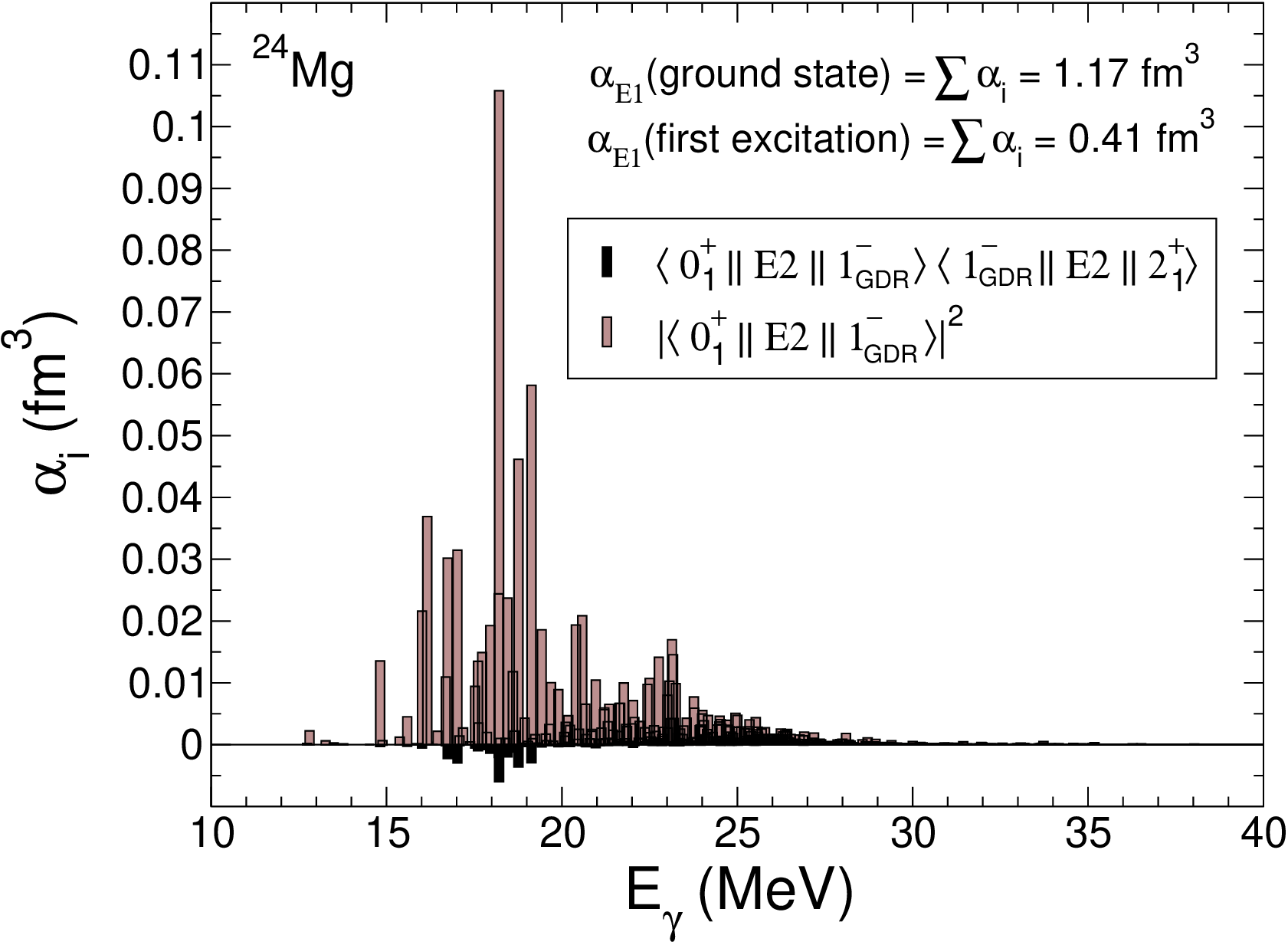}
\caption{Calculated sum of $E1$ matrix elements ($S(E1)$ in Eq.~\ref{eq:SE1}) using the shell model.}
\label{fig:finalplot}       % Give a unique label
\end{figure}

A plausible microscopic interpretation is provided by the slight change in the nuclear dipole polarizability $\alpha_{_{E1}}$ for ground and first excited (2$^+_1$) states,
as provided by second-order perturbation theory  and the hydrodynamic model~\citep{Orce2024},
\begin{eqnarray}
\alpha_{_{E1}}(\mbox{ground state}) &=& \frac{2e^2}{2J_{_i}+1}~\sum_n \frac{\big| \langle i\parallel\hat{E1}\parallel n\rangle \big|  ^2}{E_{_{\gamma}}}, \label{eq:alpha1}\\
\alpha_{_{E1}}(2^+_1) &=& 4.295\frac{Z A^{2/3} S(E1)}{\langle 0^+_1\mid\mid \hat{E2} \mid\mid 2^+_1\rangle} ~~\mbox{fm$^3$}, \label{eq:alpha2}
% {\langle i||\hat{M}(E2)||f\rangle} ~~\mbox{fm$^2/$MeV}, \\
%  \kappa(2^+_1) &=& \frac{1}{0.00039\frac{A}{Z}}\frac{S(E1)}{\langle 0^+_1\mid\mid \hat{E2} \mid\mid 2^+_1\rangle},
\end{eqnarray}for the two-step process, $0_{_1}^+ \rightarrow 1_{_{GDR}}^-\rightarrow 2_{_1}^+$, where $S(E1)$ in Eq.~\ref{eq:alpha2} is the sum  over all intermediate states
$|J_n\rangle = \large|1^-_{_{GDR}}\rangle$ connecting both $\mid 0^+_{_1}\rangle$ and $\mid 2^+_{_1}\rangle$ states via isovector $E1$ transitions following the isospin selection rule $\Delta T = 1$
for $N=Z$ self-conjugate nuclei~\citep{isospinselection},  and defined by
\begin{equation}
 S(E1):= \frac{1}{2J_i+1}
 \sum_{J_n,\Delta T} W_{inf} \frac{\langle i\parallel\hat{E1}\parallel n\rangle \langle n\parallel\hat{E1}\parallel f\rangle}{E_{_{\gamma}}},
 \label{eq:SE1}
\end{equation}
in units of e$^2$fm$^2/$MeV,  where $W_{inf}=W(1102,21)$ is the corresponding Racah W-coefficient~\citep{racah1942theory}.

The difference between $\alpha_{_{E1}}$ for the ground and first excited (2$^+_1$) states is shown in Fig.~\ref{fig:finalplot},
which presents 1$\hbar\omega$ shell-model calculations of $E1$ matrix elements using the phenomenological {\sc WBP} interaction~\citep{warburton1992effective} in the $psdpf$ model space,
as described in detail in Refs.~\citep{orce2023global,Orce2024}. It is clear that whereas the sum of matrix elements add up constructively for Eq.~\ref{eq:alpha1} yielding
$\alpha_{_{E1}}(\mbox{ground state})=1.17$ fm$^3$,
the product of $\langle i\parallel\hat{E1}\parallel n\rangle \langle n\parallel\hat{E1}\parallel f\rangle$ matrix elements may contribute destructively to $S(E1)$ in Eq.~\ref{eq:alpha2},
yielding a smaller $\alpha_{_{E1}}(\mbox{first excitation})=0.41$ fm$^3$. This results in a larger $a_{sym}$ coefficient for the first excitation as the symmetry energy is inversely
proportional to  $\alpha_{_{E1}}$~\citep{migdal1945quadrupole,levinger1957migdal}. Although these calculations are carried out at $T\approx 0$ MeV, Eq.~\ref{eq:alpha2} is, in principle,
valid
for excited states at $T\approx0.5$ MeV, where the product of $E1$ matrix elements may also contribute destructively. Unfortunately, such 1$\hbar\omega$ shell-model calculations are not feasible
for such high-excitation energies and for the heavy neutron-rich nuclei where the r-process occurs.

\begin{figure}
% Use the relevant command for your figure-insertion program
% to insert the figure file.
\centering
% \sidecaption
\includegraphics[width=8.cm,height=6.5cm,clip]{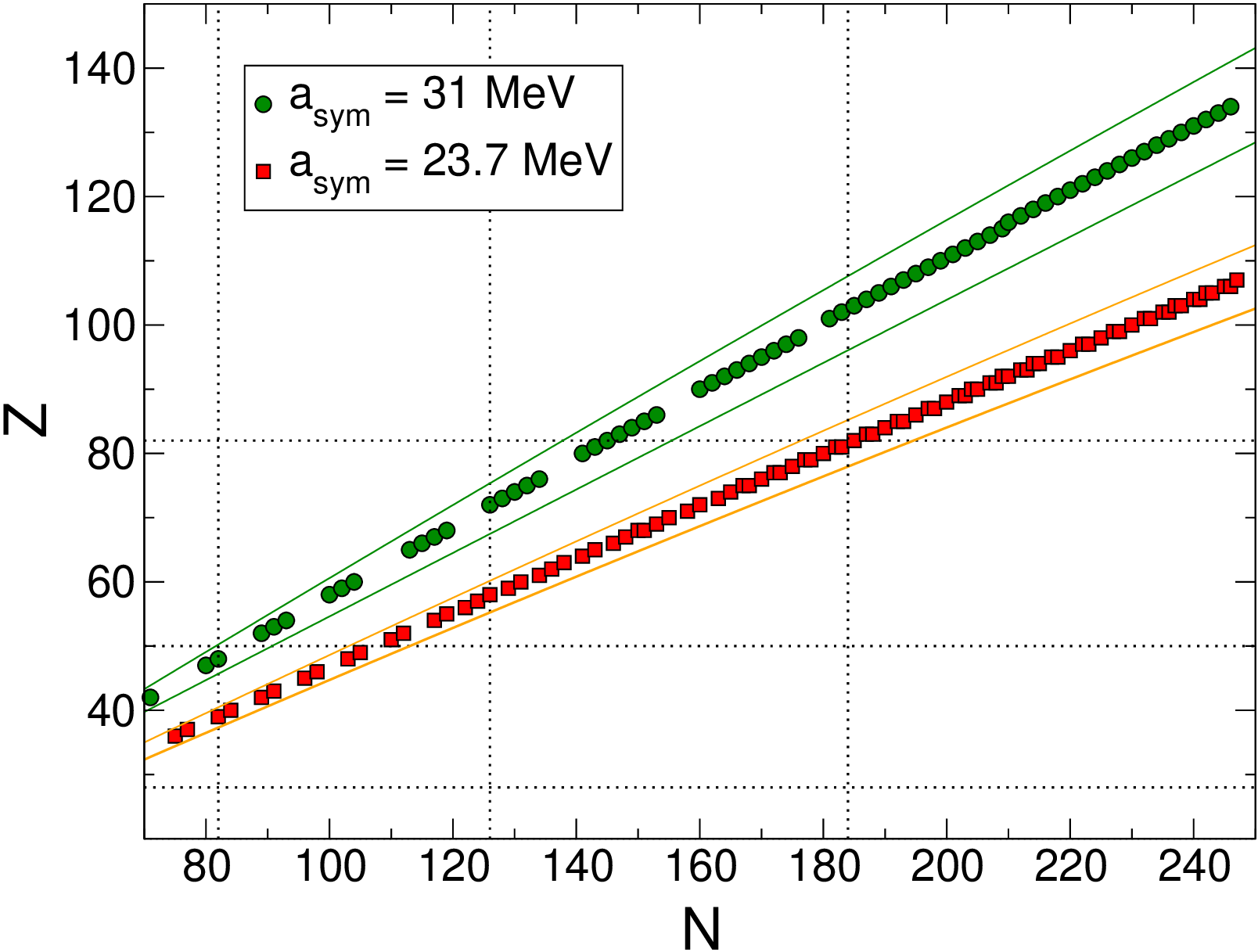}
\caption{Neutron drip lines predicted at symmetry energy coefficients of $a_{sym}=23.7$ (squares) and $31$ (circles) MeV. Solid lines indicate the loci limits determined from the
uncertainty in the symmetry energy. Dotted lines indicate the proton and neutron magic numbers..}
\label{fig:dripline}       % Give a unique label
\end{figure}

Accordingly, $a_{sym}(A)$ remains almost constant at $T\approx0.5$ MeV, the temperature where heavy elements are expected to be created through successive neutron captures
in the r-process~\citep{goriely2011r}, following the trend presented in previous work~\citep{orce2023enhanced}.
Such an increase in the symmetry energy leads to the decrease of binding energy and reduction of radiative neutron capture rates as neutron-rich nuclei become less bound~\citep{orce2023enhanced}.
The effect from a larger symmetry energy at $T\approx0.5-1$ MeV is llustrated in Fig.~\ref{fig:dripline},
which shows the corresponding neutron drip lines using the usual $a_{sym}=23.7$ MeV~\citep{rohlf1994wiley} (squares) and $31$ MeV (circles), respectively.
The nuclear chart determined using $a_{sym}=31$ MeV illustrates a substantial close-in of the neutron drip line,
% as a result of the decreasing binding energy per nucleon in neutron-rich nuclei.
% For instance, the drip line closes in from $^{254}$Pt to $^{216}$Pt for $a_{sym}=23.7$ and $31$ MeV, respectively.
% This results in a substantial close in of the neutron drip line,
as originally suggested by Berman~\citep{berman1973international} and Goriely~\citep{goriely2003r},
constraining exotic $r$-process paths far away from the line of stability and explaining the origin of the universality of heavy-element abundances through the $r$-process;
as inferred from the similar abundances observed in extremely metal-poor stars and the Sun~\citep{frebel2018nuclei,sneden2008neutron,ji2016r,kajino2019current}.
Further information on the symmetry energy  at $T\lessapprox0.5$ MeV is crucial to set the limits of nuclear existence.

% Figure \ref{fig:dripline} shows the respective neutron drip lines and clearly illustrates  the effect of an
% The larger symmetry energy in the production of heavy elements constrains exotic $r$-process paths far away from the line of stability, and
% may elucidate the
% long-sought universality of heavy-element abundances through the $r$-process; as inferred from the similar abundances observed in extremely metal-poor stars and the Sun~\citep{frebel2018nuclei,sneden2008neutron,ji2016r,kajino2019current}. Further information on the symmetry energy  at $T<0.7$ MeV
% is crucial to set the limits of nuclear existence.

% Further work at $T\lesssim0.5$ MeV is crucially required to determine the symmetry energy of heavy, neutron-rich and exotic nuclei.
% Modern radioactive-ion-beam facilities may not reach the neutron drip line, but do we really know where it lies?

% For tables use syntax in table~\ref{tab-1}.
% \begin{table}
% \centering
% \caption{Please write your table caption here}
% \label{tab-1}       % Give a unique label
% % For LaTeX tables you can use
% \begin{tabular}{lll}
% \hline
% first & second & third  \\\hline
% number & number & number \\
% number & number & number \\
% number & number & number \\\hline
% \end{tabular}
% % Or use
% \vspace*{5cm}  % with the correct table height
% \end{table}

\section*{Acknowledgement}
The author is indebted to the authors of the original work published in MNRAS,  Cebo Ngwetsheni, Balaram Dey, Deepak Pandit and Srijit Bhattacharya,
without whom this synergy between nuclear physics and astrophysics could not have been uncovered.\\

\section*{Data availability}

The data underlying this article are available from online sources at the nuclear reaction databases:
EXFOR: Experimental Nuclear Reaction Data, https://www-nds.iaea.org/exfor/exfor.htm, and ENDF: Evaluated Nuclear Data File, https://www-nds.iaea.org/exfor/endf.htm and Atomic Masses at AMDC: https://www-nds.iaea.org/amdc/

%%%%%%%%%%%%%%%%%%%% REFERENCES %%%%%%%%%%%%%%%%%%

% The best way to enter references is to use BibTeX:

\bibliographystyle{mnras}
\bibliography{enhanced}

% Don't change these lines
\bsp	% typesetting comment
\label{lastpage}
\end{document}